\documentclass[superscriptaddress,twocolumn,aps,pra,showpacs,
fixfloats]{revtex4}
\usepackage{bm}
\usepackage{dcolumn}
\usepackage{graphicx}
\begin{document}
 \title{On confinement resonances in $A$@C$_{60}$ photoionization: easy to observe?}
\author{V. K. \surname{Dolmatov}}
\email{vkdolmatov@una.edu} \affiliation{Department of Physics and
Earth Science, University of North Alabama,
Florence, Alabama 35632, USA}
\author{S. T. \surname{Manson}}
\email{smanson@gsu.edu}
\affiliation{Department of Physics and Astronomy, Georgia State
University, Atlanta, Georgia 30303, USA}
\date{\today}
\begin{abstract}
Possible reasons that confinement resonances are not observed in a
recent photoionization experiment on the endohedral fullerene Ce@C$_{82}^{+}$ are explored. The effect of the polarization of the fullerene
shell in response to the ionization of the endohedrally encaged atom $A$@C$_{n}$, termed the ``shielding'' effect, has been investigated and
found to be relatively
small; no more than a $20\%$ effect near threshold, and much less at higher energies. It is argued that most likely,
the absence of confinement resonances in Ce@C$_{82}^{+}$ is due primarily to the finite thickness of the carbon cage; the
off-the-center position and thermal vibration of the encaged atom, discussed elsewhere, further weaken the resonances rendering them beyond
the sensitivity of the experiment to detect, in this case.  For other
situations/endohedrals, the confinement resonances should well be observable, and Ne@C$_{60}$ is suggested as an excellent candidate.
\end{abstract}
\maketitle

Photoionization spectra of atoms $A$ encapsulated inside hollow
fullerene cages C$_{n}$, referred to as endohedral atoms $A$@C$_{n}$, have been the subject of extensive
theoretical study by a number of investigators at various levels of approximation for about a decade or so.
 The significance of these studies is that they provide information on how confinement changes properties of atoms.
This information, in turn, is of considerable importance to basic science, as well as to applied sciences and
technologies, since confined atoms are of multidisciplinary significance, see, e.g., Ref.~ \cite{Connerade}.

An outstanding feature of the photoionization spectrum of the endohedral atom $A$@C$_{n}$  is associated with predicted resonances,
termed confinement resonances, in the spectrum that have been extensively studied theoretically to date.
The resonances occur due to the interference of the
photoelectron waves emerging directly from the encapsulated (confined) atom $A$ and those scattered off the
C$_{n}$ confining cage \cite{DolmatovAQC,HimadriPRA'10,Korol'10,Ludlow'10,BaltenkovJPB99,ConneradeJPB99,ConneradeJPB00}.
Note that oscillations of the similar nature have been seen, both theoretically and experimentally, in the photoionization of the free C$_{60}$
fullerene where
the photoelectron emerges from the C$_{60}$ itself \cite{MMC}. This phenomenology provides strong supporting evidence for the predicted existence of
confinement resonances arising from the photoionization of the encaged atom in $A$@C$_{60}$ as well.

It has been not until very recently that a reliable experiment on the photoionization spectrum of
an endohedral atom $A$@C$_{n}$ has become possible \cite{MoellerPRL08}. There, the $4d$
photoabsorption spectrum of Ce in Ce@C$_{82}^{+}$ has been experimentally measured. Aside from a very intriguing
and still unresolved discovery of a significant redistribution of Ce $4d$ oscillator strengths in Ce@C$_{82}^{+}$ compared to
free Ce, another finding was that no confinement resonances were observed in the spectrum, in contrast to expectations.
Such expectations were primarily driven by theoretical results \cite{AmusiaJPB05} where
very strong confinement resonances in the $4d$ photoionization spectrum of Xe in Xe@C$_{60}$ were predicted. Note, in all cited theoretical studies associated
with confinement resonances, including Ref. \cite{AmusiaJPB05} on Xe@C$_{60}$, the encapsulated atom
was positioned at the center of the cage.

The experimental data on Ce@C$_{82}^{+}$, on the one hand, challenge one to explain the absence of traces of confinement resonances
in the Ce@C$_{82}^{+}$ spectrum. A suggested reason for this is associated with the off-the-center
position of Ce in Ce@C$_{82}^{+}$ \cite{Korol'10,BMM} along with its thermal vibrations \cite{Korol'10}. In the present paper, however,
it is pointed out that confinement
resonances in the $4d$ absorption spectrum of Ce@C$_{82}^{+}$ could hardly be observable even if the Ce atom were at the center of
the cage, i.e., even omitting the off-the-center and thermal vibrations effects.
On the other hand, the experiment poses a more troubling question, namely, whether some effects have been overlooked in the original
theoretical predictions of confinement resonances that
might ``wash out'' the resonances in the photoionization of an \textit{at-the-center} encapsulated atom $A$ in a C$_{n}$ fullerene.

In any of earlier theoretical work on confinement resonances, the ``shielding'' effect was not considered in detail. The quintessence of the shielding effect
is that while an outgoing photoelectron  is passing
through the C$_{n}$'s wall, it could be partially or totally shielded from the Coulomb field of the final-state ion $A^{+}$ when it is
between the inner and outer surfaces of the carbon cage. This is because the final-state ion
$A^{+}$ can polarize the  fullerene cage similar to a conducting shell. The electric potential
of the central $A^{+}$ is totally or partially canceled out by the electrostatic potential of an induced negative charge on the inner surface of C$_{n}$,
thereby changing the potential experienced by the photoelectron while it is between the inner and outer surfaces of the
C$_{n}$. The effect of total cancelation will occur if the cage acts as a perfect conductor. A partial cancelation will ensue if the cage
is only somewhat polarizable.  The impact of the shielding effect on confinement resonances is detailed below.

Following our earlier work \cite{DolmatovAQC,ConneradeJPB99,ConneradeJPB00}, a spherical C$_{n}$ cage will be modeled by a
 short-range, spherical potential $U_{\rm c}(r)$ of the inner radius $r_{0}$, thickness $\Delta$, and depth $U_{0}$:
\begin{eqnarray}
 U_{\rm c}(r)=\left\{\matrix {
U_{0} <0, & \mbox{if $r_{0} \le r \le r_{0}+\Delta$} \nonumber \\
0 & \mbox{otherwise.} } \right.
\label{eqU}
\end{eqnarray}
For C$_{60}$, $r_{0}= 5.8$ a.u., $\Delta=1.9$ a.u., and $U_{0}= -8.2$ eV \cite{DolmatovAQC,ConneradeJPB99,ConneradeJPB00}.
The applicability of the model is generally limited to low
photoelectron energies when the photoelectron wavelength significantly exceeds the bond length between the carbon atoms
of C$_{n}$, so that the C$_{n}$ cage ``looks'' as a homogeneous charge distribution to the outgoing photoelectron.
This potential is added to the atomic Hartree-Fock Hamiltonian $\hat{H}_{0}^{\rm HF}$ (which is defined in the manner of an isolated atom)
 thereby forming a ``confined'' HF Hamiltonian
 \begin{eqnarray}
 \hat{H}_{\rm c}= \hat{H}_{0}^{\rm HF}+U(r).
 \label{eqH}
 \end{eqnarray}
  Solutions of the corresponding
 ``confined'' HF equation $\hat{H}_{\rm c}\psi_{i}(\bm{r})= E_{i}\psi_{i}(\bm{r})$ give one the initial
 ground-state electronic energies $E_{i}$ as well as both the ground-state $\psi_{i}(\bm{r})$ and final-state $\psi_{f}(\bm{r})$ wavefunctions.
 This is how the energies and wavefunctions of a confined atom were defined originally \cite{DolmatovAQC,ConneradeJPB99,ConneradeJPB00}.

To account for the shielding effect on an outgoing photoelectron while it is passing through the C$_{n}$, we introduce a re-defined
  final-state ``confined'' Hamiltonian
$\hat{{\cal{H}}}_{c}$ which replaces $\hat{H}_{\rm c}$ in the HF equation,
\begin{eqnarray}
\hat{{\cal{H}}}_{\rm c}=\left\{\matrix {
\hat{H}_{\rm c} - \frac{\alpha}{r}, & \mbox{if $r_{0} \le r \le r_{0}+\Delta$} \nonumber \\
\hat{H}_{\rm c}, & \mbox{otherwise.} } \right.
\label{eqHF^f}
\end{eqnarray}
Here, $\alpha =0$ if the shielding effect is ignored, as in earlier studies, $\alpha = 1$ for
complete shielding of the ion-core potential of the final-state ion $A^{+}$ by the polarized C$_{n}$ cage between the inner and outer
wall of the C$_{n}$, and $\alpha < 1$ if only partial shielding of the $A^{+}$'s potential occurs. Finally, when solving the corresponding
HF equations for the final-state
wavefunctions $\psi_{f}(\bm{r})$ of the outgoing photoelectron, the latter are orthogonalized to the ground-state wavefunctions
$\psi_{i}(\bm{r})$ with using Lagrange's off-diagonal parameters $\lambda_{ij}$ in the exactly same manner as for an
isolated atom \cite{AmusiaATOM}. This corrects for the use of different potentials in the ground-state
and final-state HF calculations. The thus defined electronic energies and wavefunctions are starting points for the random phase approximation with
exchange (RPAE) equation \cite{AmusiaATOM} which allows one to calculate photoionization matrix elements, their phase shifts, and
total and differential photoionization cross sections of the $A$@C$_{n}$ atom.

\begin{figure}[h] \includegraphics[width=7cm]{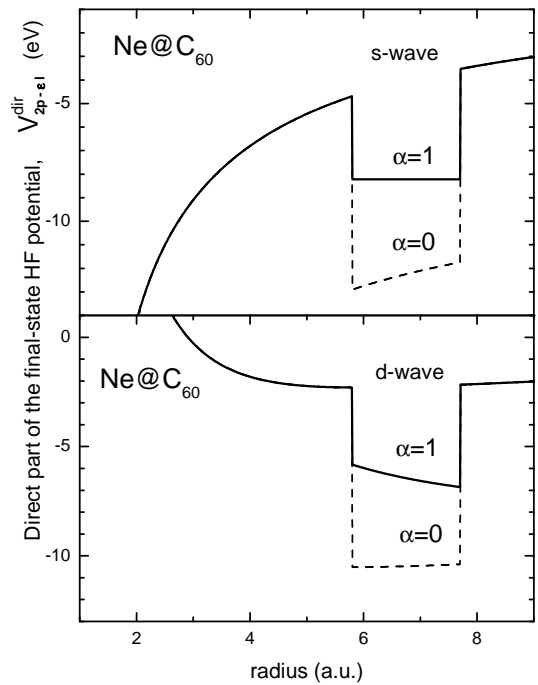}
\caption{ ${\cal{V}}_{nl \rightarrow \epsilon l'}^{\rm dir}(r)$ potential seen by the $\epsilon s$
and $\epsilon d$ photoelectron waves upon the $2p$ photoionization of Ne@C$_{60}$ calculated both with ($\alpha =1$)
and without ($\alpha = 0$) accounting for
a complete shielding of the direct atomic
potential of Ne$^{+}$ by the static potential of the polarized C$_{60}$, as discussed in the body of the paper.}
\label{fig1}
\end{figure}

To begin with, the direct part of the potential ${\cal{V}}_{nl -\epsilon l'}^{\rm dir}(r)$ of the final-state Hamiltonian $\hat{{\cal{H}}}_{\rm c}$
seen by an outgoing $\epsilon l'$ photoelectron due to the photoionization of $A$@C$_{n}$ is given by
\begin{eqnarray}
{\cal{V}}_{nl \rightarrow \epsilon l'}^{\rm dir} = V_{nl \rightarrow \epsilon l'}^{\rm dir}(r) + \frac{l'(l'+1)}{2r^{2}} +U_{\rm c}(r)-\frac{\alpha}{r}.
\label{Vdir}
\end{eqnarray}
Here, $V_{nl \rightarrow \epsilon l'}^{\rm dir}(r)$ is a direct atomic Hartree-Fock potential with the excluded centrifugal potential. Note,
the atomic exchange potential can
be safely ignored at the C$_{n}$
boundary since the radius of C$_{n}$ considerably exceeds the atomic size. As an example, the calculated
${\cal{V}}_{nl \rightarrow \epsilon l'}^{\rm dir}$ potentials seen by the $\epsilon s$
and $\epsilon d$ photoelectron waves upon the $2p$ photoionization of Ne@C$_{60}$ are shown in Fig.~\ref{fig1}.
One can see that the shielding effect noticeably reduces the potential depth of
${\cal{V}}_{nl \rightarrow \epsilon l'}^{\rm dir}$ inside the C$_{60}$. Thus, one might
expect a noticeable decrease in the confinement resonances, since the coefficient of
reflection of the outgoing photoelectron wave by the potential well lessens with decreasing depth
of the well.

\begin{figure} \includegraphics[width=7cm]{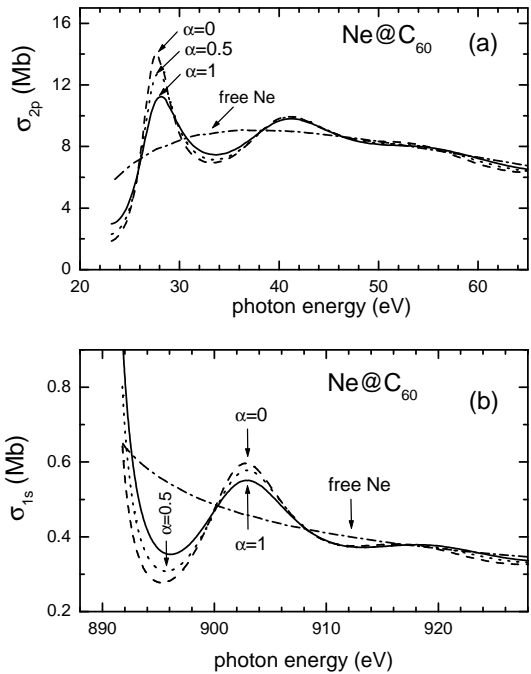}
\caption{RPAE calculations of the photoionization cross sections $\sigma_{1s}(\omega)$ and $\sigma_{2p}(\omega)$
of confined Ne (Ne@C$_{60}$) and free Ne, as marked. Dashed line, calculations without  accounting for
shielding of the direct atomic potential of Ne$^{+}$ by the static potential of the polarized C$_{60}$
($\alpha = 0$). Dotted line, calculations accounting for only partial shielding ($\alpha = 0.5$).
Solid line, calculations accounting for complete shielding ($\alpha =1$). }
\label{fig2}
\end{figure}

To illustrate the actual changes in the strengths of confinement resonances, the RPAE calculated
$1s$  and $2p$  photoionization cross sections [$\sigma_{1s}(\omega)$ and $\sigma_{2p}(\omega)$, respectively]
of
Ne@C$_{60}$ versus the photon energy $\omega$ for various values of the shielding parameter $\alpha$ are depicted in Fig.~\ref{fig2}.
One can see
that even complete shielding ($\alpha = 1$) results in a no more than a $20\%$ change in the photoionization
cross sections near threshold (compared to a ``non-shielded'' result, $\alpha=0$), and much less at higher photon energies.
Note, the value of the parameter $\alpha$ for actual shielding is, obviously, somewhere between the extreme values of $\alpha =0$ and $\alpha =1$.
Hence, the actual impact of the shielding effect on photoionization cross sections is yet smaller than $20\%$ (cf., e.g.,
results for $\alpha =0.5$). One can thus conclude that the shielding effect is relatively insignificant for $A$@C$_{n}$
photoionization. Thus, it can safely be excluded as a possible factor for ``washing out'' confinement resonances from
the photoionization spectrum of $A$@C$_{n}$.

\begin{figure}[h] \includegraphics[width=8cm]{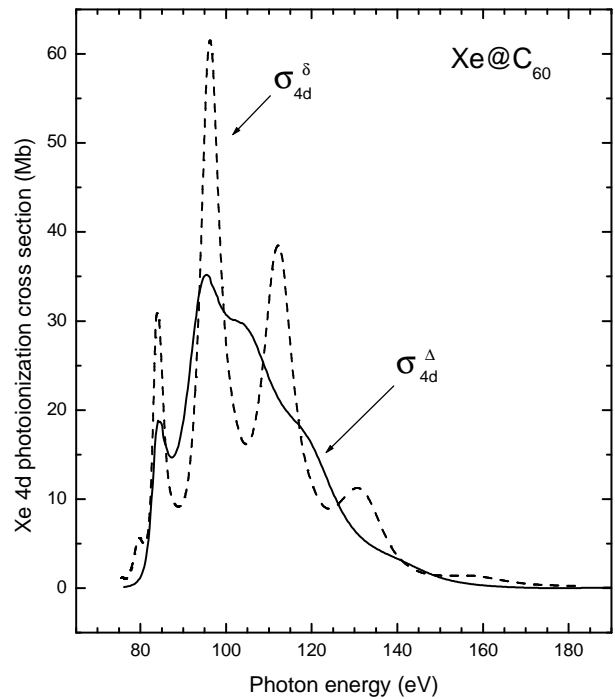}
\caption{
RPAE results for the $4d$ photoionization cross of confined Xe
(Xe@C$_{60}$) calculated within the framework of both the zero-thickness $\delta$-potential
model, $\sigma_{4d}^{\delta}$ \cite{AmusiaJPB05}, and accounting for the finite-thickness $\Delta$ of C$_{60}$, $\sigma_{4d}^{\Delta}$
\cite{DolmatovXe4d}. The spikes in the cross section are confinement resonances.}
\label{fig3}
\end{figure}

Then, what could be possible reasons for the absence of confinement resonances in the experimental $4d$
photoabsorption spectrum of Ce@C$_{82}^{+}$? Some reasons for this, like the off-center position and thermal vibrations of the confined ion,
were proposed in Refs.\ \cite{Korol'10,BMM}.
However, we believe that the primary reason for the absence of confinement resonances in the Ce@C$_{82}^{+}$ spectrum
is of a different nature. First of all, the expectations of seeing strong resonances in the Ce@C$_{82}^{+}$ spectrum were
based on the predictions made for the Xe $4d$ absorption spectrum of Xe@C$_{60}$ in Ref.\ \cite{AmusiaJPB05} where
the ``delta''-potential model was used for the description of photoionization
of $A$@C$_{60}$ atoms. In this modeling,  the C$_{60}$ cage is modeled by a spherical
Dirac bubble potential $U(r) =-A\delta(r-r_{0})$, where $r_{0}$ is the radius from the center of the C$_{60}$
cage to the middle point between the inner and outer surfaces of the cage, and $A$ is
the potential strength.  The $\delta$-potential model
thus assumes that the C$_{60}$ cage has a zero thickness, $\Delta=0$. The predicted confinement resonances on the basis of this model,
including confinement resonances in the Xe $4d$ photoionization of Xe@C$_{60}$, are huge.  It is not surprising that
researchers were expecting the same strong resonances to occur in the $4d$ photoabsorption spectrum of confined Ce as well.
However, as was demonstrated in Ref.\ \cite{DolmatovXe4d}, and for a number of other examples in Ref.\ \cite{DolmatovAQC},
accounting for a finite thickness $\Delta$ of the C$_{60}$ cage lessens the strength of confinement resonances considerably. This is illustrated
in Fig.~\ref{fig3} where the ``delta''-potential model calculated results for the $4d$ photoionization of Xe@C$_{60}$ \cite{AmusiaJPB05}
are compared with those \cite{DolmatovXe4d} obtained in the  finite-C$_{60}$-thickness potential model [Eqs.\ (\ref{eqU}) and (\ref{eqH})].

 One can see that accounting for finite
 thickness $\Delta$ dramatically reduces the confinement resonances in the $4d$ spectrum of Xe@C$_{60}$. We believe that
 the same is true for the $4d$ spectrum of Ce@C$_{82}^{+}$ as well.
 Such weak resonances are probably not observable yet experimentally, owing to the difficulty of performing experiments with such a small
number of molecules as are currently available.

In conclusion, it is shown that the possible effects of shielding of the $A^{+}$'s field, owing to the polarization of confining shell,
are quite weak and, thus, can not lead to the disappearance of the confinement resonances from the spectrum of an endohedral
atom $A$@C$_{n}$. As for the confinement resonances in the $4d$ spectrum of Ce@C$_{82}^{+}$, it is argued (on the basis of an
analogy with the $4d$ spectrum of Xe@C$_{60}$) that
 they must be quite weak due to the finite-thickness effect. Additionally, the off-the-center-position and
 thermal vibrations effects further suppress the strength of the resonances. Hence, the
 combination of the finite-thickness, off-the-center position, and thermal vibrations effects make the confinement resonances
 in the Ce@C$_{82}^{+}$ spectrum so insignificant that they, while being there in principle, are not observable experimentally, contrary
to expectations based on a zero-thickness model of the C$_{60}$ potential, a model that seems to be quantitatively inaccurate in this situation.
These considerations explain the absence of confinement resonances in the measured $4d$ spectrum of Ce@C$_{82}^{+}$. In other situations,
 particularly for at-the-center confined atoms, the confinement resonances are likely strong and observable, e.g., see
Fig.~\ref{fig2} for the $1s$ and $2p$ near threshold photoionization of Ne@C$_{60}$. We suggest the latter as a test case for experimental
study.

This work was supported by NSF and DOE, Basic Energy Sciences. The authors thank Emre Guler for the help in performing the calculations.


\begin{thebibliography}{}
%
\bibitem{Connerade} J.-P. Connerade, in \textit{The Fourth International Symposium on Atomic Cluster Collisions},
edited by A. V. Solv'yev and E. Surdutovich, AIP Conference Proceedings \textbf{1197}, 1 (2009).
%
\bibitem{DolmatovAQC} V. K. Dolmatov, in \textit{Theory of Confined Quantum Sytems: Part
Two}, edited by J. R. Sabin and E. Br\"{a}ndas, Advances in Quantum
Chemistry (Academic Press, New York, 2009), Vol. 58,
pp. 13-68.
%
\bibitem{HimadriPRA'10} M. E. Madjet, T. Renger, D. E. Hopper, M.A. McCune, H. S. Chakraborty,
J.-M. Rost, and S. T. Manson, Phys. Rev. A \textbf{81}, $013202$ ($2010$).
%
\bibitem{Korol'10} A. V. Korol and A. V. Solov'yov, arXiv: 0912.2690v2 (2009).
%
\bibitem{Ludlow'10} J. A. Ludlow, T.-G. Lee, and M. S. Pindzola, Phys. Rev. A \textbf{81}, 023407 (2010).
%
\bibitem{BaltenkovJPB99} A. S. Baltenkov, J. Phys. B \textbf{32}, 2745 (1999).
%
\bibitem{ConneradeJPB99} J. P. Connerade, V. K. Dolmatov, and S. T. Manson, J. Phys. B \textbf{32}, L395 (1999).
%
\bibitem{ConneradeJPB00} J. P. Connerade, V. K. Dolmatov, and S. T. Manson, J. Phys. B \textbf{33}, 2279 (2000).
%
\bibitem{MMC} M. A. McCune, M. E. Madjet and H. S. Chakraborty, J. Phys. B \textbf{41}, 201003 (2008) and references therein.
%
\bibitem{MoellerPRL08} A. M\"{u}ller, S. Schippers, M. Habibi, D. Esteves, J. C. Wang, R. A. Phaneuf, A. L. D. Kilcoyne,
A. Aguilar, and L. Dunsch, Phys. Rev. Lett. \textbf{101}, 133001 (2008).
%
\bibitem{AmusiaJPB05} M. Ya. Amusia, A. S. Baltenkov, L. V. Chernysheva, Z. Felfli, and A. Z. Msezane, J. Phys. B
\textbf{38}, L169 (2005).
%
\bibitem{BMM} A.S. Baltenkov, S. T. Manson and A. Z. Msezane, J. Phys. B (submitted).
%
\bibitem{AmusiaATOM} M. Ya. Amusia and L. V. Chernysheva, \textit{Computation of Atomic
Processes} (IOP Publishing Ltd., Bristol, 1997).
%
\bibitem{DolmatovXe4d} V. K. Dolmatov and S. T. Manson, Phys. Rev. A \textbf{41}, 165001 (2008).
\end{thebibliography}
\end{document}